\documentclass[aps,pra,twocolumn,showpacs]{revtex4-1}
\usepackage{amsbsy,latexsym,amsmath}
\usepackage{amsfonts}
\usepackage{amssymb}
\usepackage[mathscr]{eucal}
\usepackage{epsfig,graphics,graphicx}
\usepackage{color}

\begin{document}

 \title{Optimal Cloning of Quantum States with a Fixed Failure Rate}
\author{E. Bagan$^{1,2}$, V. Yerokhin$^{1}$, A. Shehu$^{1}$, E. Feldman$^{3}$, and J. A. Bergou$^{1}$}
\affiliation{$^{1}$Department of Physics and Astronomy, Hunter College of the City University of New York, 695 Park Avenue, New York, NY 10065, USA\\
$^{2}$F\'{i}sica Te\`{o}rica: Informaci\'{o} i Fen\`{o}mens Qu\`antics, Universitat Aut\`{o}noma de Barcelona, 08193 Bellaterra (Barcelona), Spain \\
$^{3}$Department of Mathematics, Graduate Center of the City University of New York, 365 Fifth Avenue, New York, New York 10016, USA}

\begin{abstract} 
Perfect cloning of a known set of states with arbitrary prior probabilities is possible if we allow the cloner to sometimes fail completely. In the optimal case the probability of failure is at its minimum allowed by the laws of quantum mechanics. Here we show that it is possible to lower the failure rate below that of the perfect probabilistic cloner but the price to pay is that the clones are not perfect; the global fidelity is less than one. We determine the optimal fidelity of a cloner 
with a Fixed Failure Rate (FFR cloner) in the case of a pair of known states. Optimality is shown to be attainable by a measure-and-prepare protocol in the limit of infinitely many clones. The optimal protocol consists of discrimination with a fixed rate of inconclusive outcome followed by preparation of the appropriate clones. The convergence shows a symmetry-breaking second-order phase transition in the fidelity of the approximate infinite clones.

\end{abstract}
\maketitle 

Probabilistic protocols enable us to carry out tasks that according to the laws of quantum mechanics are impossible deterministically, leading to important applications for quantum information processing. Remarkable examples are unambiguous state discrimination, whereby non-orthogonal states can be identified without error~\mbox{\cite{Ivanovic,Dieks,Peres,Jaeger&Shimony}}, and perfect cloning of a known set of states, which can be performed probabilistically, although its deterministic version is forbidden by the no-cloning theorem. More recent advances include replication~\cite{Chiribella}, quantum state amplification~\cite{amplification} (and references therein), and probabilistic metrology~\cite{prob metr}.

The price to pay for making the impossible possible is to allow the protocols to fail sometimes. To assess the amount of resources that a given task will require one must know what is the failure probability $Q$ of the corresponding protocol. The minimum failure probability~$Q_{\rm min}$ or, alternatively, the maximum success probability, often defines the optimal probabilistic protocol. A relevant practical question arises at this point, particularly if $Q_{\rm min}$ is large: Can one reduce the failure probability by allowing slight deviations from the perfect output to occur? 
The answer is known to be in the affirmative for unambiguous discrimination, as shown first in~\cite{Chefles+Barnett 1st} for the case of two pure states with equal priors.  Recently, the solution for general priors was obtained in two alternative ways. Either one can introduce a fixed error probability as in~\cite{hayashi1,hayashi2} or, equivalently, one can fix the failure probability below the minimum $Q$ as in~\cite{FRIO,herzog}.  We will show that the latter bears strong connections with cloning. We refer to it as Fixed Rate of Inconclusive Outcome scheme, or FRIO scheme for short. 

Here, we address the same question for cloning. We will show that indeed approximate clones can be obtained for failure rates below the minimum failure rate $Q_{\rm min}\equiv Q_{\rm PC}$ of perfect cloning. We refer to this scheme as Fixed Failure Rate (FFR) cloning. It was first proposed by Chefles and Barnett~\cite{Chefles+Barnett inter} who considered two non-orthogonal states with equal priors. We extend their results to general priors. This is of practical relevance, as for implementations one must know the robustness of the protocol against perturbations of the various parameters involved. 

From a fundamental viewpoint, the equal prior case is too restricted to provide a full account of cloning. We show here that it misses the rich structure of the full solution. In particular, it misses the appearance of a phenomenon analogous to a second-order symmetry-breaking phase transition in the limit when infinite number of clones are produced. This phenomenon has been recently noticed in perfect cloning~\cite{us1}. Similar phase transition-like phenomena have been identified in other cloning scenarios~\cite{Chir&Yang}, where the fidelity of the optimal asymptotic clones reveals a universal behavior: its scaling depends solely on the number of free parameters needed to specify the input states, independently of any specific detail. Even more noticeably, for cloning of quantum clocks~\cite{Gendra} such universal scaling depends only on the number of incommensurable units required to specify the energy spectrum of the clocks. It hence appears that emergent behavior is a general feature of cloning, though a full understanding requires further work.

Another aspect of asymptotic cloning is that in the limit of infinitely many clones optimality is conjectured to  be attainable by a measure and prepare protocol, so that no coherent processing of the input states is required in this limit. Although there is no general proof to date, the conjecture has been shown to hold for the universal cloner~\cite{Bae,Chiribella2006}. In this scenario cloning becomes equivalent to state estimation followed by state preparation, and this has been shown to be so in most of the cloning scenarios covered in the literature~\cite{Chir&Yang}. For probabilistic perfect cloning the measure and prepare protocol, named ``cloning by discrimination" in~\cite{us1}, was proved to be optimal when infinitely many clones are produced.  In this paper we prove the conjecture for probabilistic approximate cloning by showing that the optimal FRIO discrimination measurement produces a classical output based on which a precise preparation of the approximate clones suffices to attain optimality.
It should be stressed that the solution to the general FRIO discrimination problem~\cite{hayashi2,FRIO,herzog} appeared more than a decade after it was originally proposed for equal priors in [8]. 
Even more surprisingly, the generalization of the probabilistic approximate cloning from equal priors, proposed almost two decades ago~\cite{Chefles+Barnett inter}, to arbitrary ones has been an open problem until now. In the present work we employ a geometric approach~\cite{us1,us2}, which has proved very powerful in dealing with highly nonlinear problems, to obtain the complete analytical solution in parametric form. Furthermore, we establish a connection between these two protocols that holds in general in the asymptotic limit.    

We now consider the optimal cloner in detail focusing on $1\to n$ cloning for simplicity. We assume the two states to be cloned are given with {\it a priori} probabilities~$\eta_1$ and~$\eta_2$, such that~$\eta_1+\eta_2=1$ and, without loss of generality, that they satisfy $\eta_1\le \eta_2$. For~\mbox{$m\to n$} cloning we just make the replacement $|\psi_k\rangle\to|\psi_k^m\rangle\equiv|\psi_k\rangle^{\otimes m}$, $k=1,2$.
 Then a natural cost function for a probabilistic cloner is given by the average failure probability 
\begin{equation}
Q=\eta_1 q_1+\eta_2 q_2,
\label{obj fun}
\end{equation}
where $q_k$ is the failure probability if the state $|\psi_k\rangle\in{\mathscr H}$ is fed into the cloner. 

Let $|\Psi_k\rangle\in{\mathscr H}^{\otimes n}$ be the state of the~$n$ clones of~$|\psi_k\rangle$. 
Ideally, one would like the cloner to produce perfect copies, i.e.,  $|\Psi_k\rangle=|\psi^n_k\rangle$. According to the no-cloning theorem this requires a minimum failure probability $Q_{\rm PC}>0$. 
The problem of optimal perfect cloning has been addressed and solved in full generality only very recently~\cite{us1}. Here we address the problem of the optimal imperfect cloner
for a given fixed failure rate~$Q< Q_{\rm PC}$ and derive the FFR cloner that produces the best approximate clones, i.e., attains the highest fidelity compatible with the fixed value of $Q$.

Our approach is based on the Neumark extension where a quantum device, in our case a cloner, is described by a unitary transformation $U$ acting on the input state $|\psi_k\rangle$ and some conveniently chosen ancillary system. We assume that initially the ancilla is in a reference state $|0\rangle$. $U$ transforms the system composed of the input and the ancilla into the state 
\begin{eqnarray}
U|\psi_k\rangle|0\rangle&=& \sqrt{p_k}|\Psi_k\rangle|s\rangle +\sqrt q_k |\Phi\rangle|f\rangle. \label{U2}
\end{eqnarray}
Here, $|s\rangle$ and $|f\rangle$ refer to two orthogonal states of a part of the ancillary system that play the role of a flag. By reading the state of the flag we know whether cloning has succeeded ($s$) of failed ($f$). If cloning has succeeded the output is in the approximate clone state $|\Psi_k\rangle$. If cloning has failed, the output is in a failure state $|\Phi\rangle$. Optimality requires  $|\Phi\rangle$ to be the same for both inputs~\cite{us1,us2}.

Taking inner products of Eq.~\eqref{U2} with the same, resp., different $k$ yields $p_k+q_k=1$ and the unitarity constraint,
\begin{equation}
s=\sqrt{p_1 p_2}\, s'+\sqrt{q_1 q_2},
\label{unit cond}
\end{equation}
where $s \equiv \langle\psi_1|\psi_2\rangle$ and $s' \equiv \langle\Psi_1|\Psi_2\rangle$. With no loss of generality, $s$ and $s'$ are assumed real and positive. If Eq.~(\ref{unit cond}) is satisfied, $U$ can be extended to a full unitary on the whole Hilbert space.

If the cloner is fed with the state $|\psi_k\rangle$, it delivers~$|\Psi_k\rangle$ with probability $p_k$ and fails with probability~$q_k=1-p_k$. The total failure probability is $Q=\eta_1q_1+\eta_2q_2$ and the total success probability is $1 - Q \equiv \bar Q$. The global fidelity of the clones $|\Psi_k\rangle$ is $F_k=|\langle\psi_k^n|\Psi_k\rangle|^2$. Following~\cite{Chefles+Barnett inter}, we asses the quality of the FFR cloner  via the average global \mbox{fidelity}, conditioned on successfully cloning the input state, 
\begin{equation}
F=\tilde\eta_1|\langle \psi^n_1|\Psi_1\rangle|^2 +\tilde\eta_2|\langle \psi^n_2|\Psi_2\rangle|^2,
\end{equation}
where $\tilde\eta_k\equiv \eta_kp_k/\bar Q$, with $\tilde\eta_1+\tilde\eta_2=1$, are the posterior probabilities conditioned on success. The maximum value of the fidelity is (see appendixes)
\begin{equation}
F_{\rm FFR}=\frac{\bar Q+\sqrt{\bar Q^2-4\eta_1\eta_2\zeta_{\rm min}^2}}{2\bar Q} ,
\label{Fmax}
\end{equation}
where $\zeta_{\rm min}$ is the minimum positive value of $\zeta$, defined as
\begin{eqnarray}
\zeta&=&(s-\sqrt{q_1q_2})\sqrt{1-s^{2n}}\nonumber\\
&-&s^n\sqrt{1-s^2+2s\sqrt{q_1q_2}-(q_1+q_2)}.
\label{zeta-means}
\end{eqnarray}
Note that~$\zeta$ is independent of the priors $\eta_1$ and $\eta_2$.

To find the best cloner for a given $Q$, we must optimize~$q_1$ and~$q_2$ so that~$\zeta$ is minimized.  Thus,
the original optimization problem is now cast as
$
\min_{q_1,q_2}\!\zeta$ ($\equiv \zeta_{\rm min}$), subject to $\eta_1 q_1+\eta_2 q_2=Q$, $\zeta\ge0$, and $0\le q_1,q_2\le 1
$.

\begin{figure}[h]
$$
\includegraphics[width=26em]{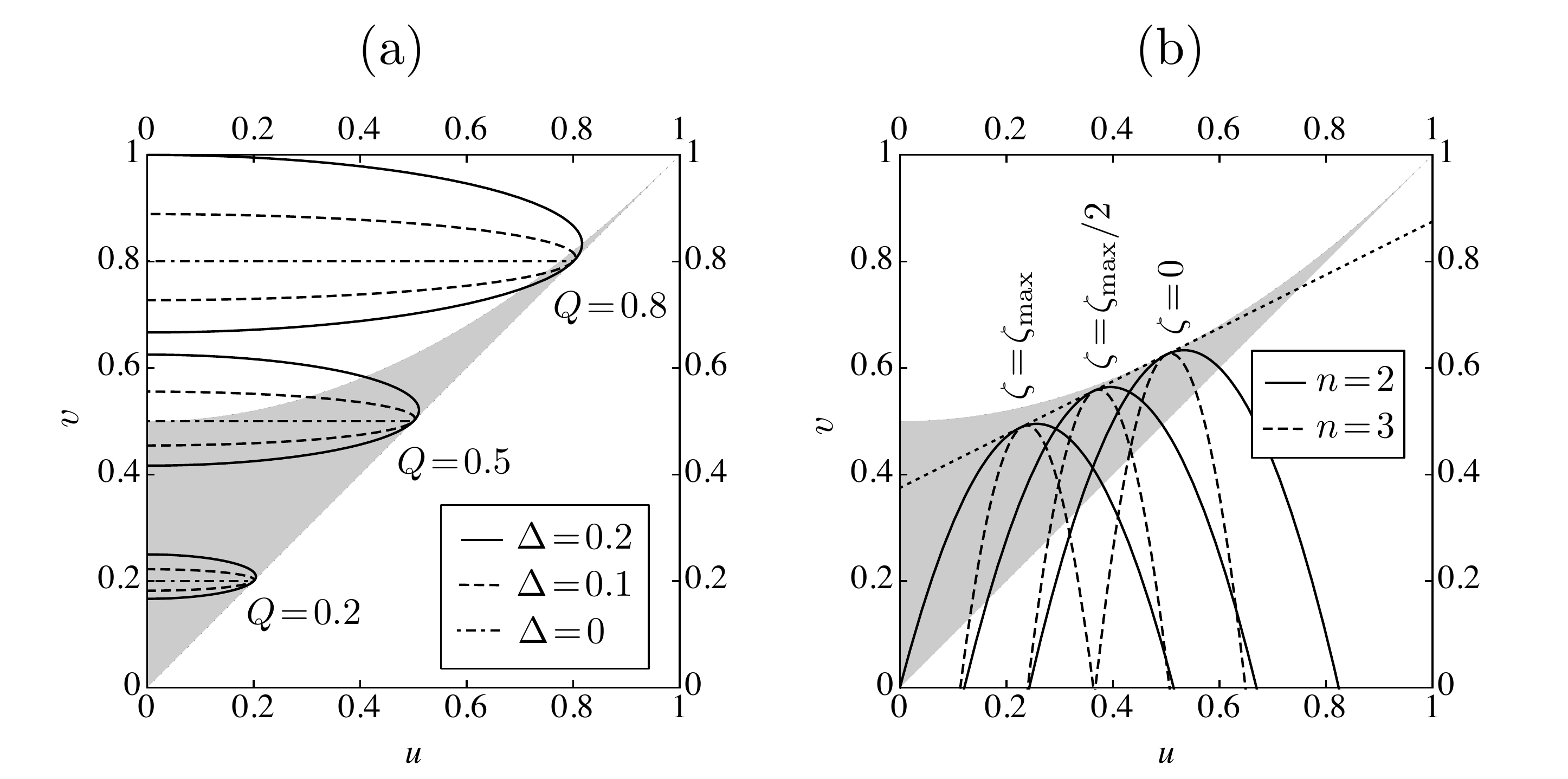}
$$
\vspace{-2em}
\caption{(a) The ellipses in Eq.~(\ref{obj fun conic}) for various values of $Q$ and $\Delta$. The geometric solution to optimal cloning falls in the gray region, whose bottom boundary, the line~$v=u$, is the envelope of these families of ellipses.  (b) Families of parabolas defined by Eq.~(\ref{para}) for fixed $s$ (in the figure~$s=0.5$) and various values of $\zeta$ and $n$. The value of $\zeta_{\rm max}$, shown for the solid-line family, is~$\zeta_{\rm max}=0.268$. The dotted straight line is the envelope of these families of parabolas, obtained by varying~$\zeta$. The top boundary of the gray region, given by $v=(1+u^2)/2$, is the envelope of this family of straight lines, obtained by varying~$s$.}
\label{fig:const z}
\end{figure}

We next develop a complete geometric approach to this optimization problem which delivers a full analytic solution in parametric form. Eq.~(\ref{zeta-means}) is a function of the arithmetic and geometric means of the failure probabilities $q_1$ and $q_2$ alone. So, let us define $u=\sqrt{q_1q_2}$, $v=(q_1+q_2)/2$.
%
We recall from~\cite{us2} that the map $(q_1,q_2)\mapsto(u,v)$ turns the straight line,~Eq.~(\ref{obj fun}), into the ellipse
\begin{equation}
u\!=\!\frac{Q}{\sqrt{1\!-\!\Delta^2}}\cos\phi,\quad
v\!=\!\frac{Q}{1\!-\!\Delta^2}\!+\!\frac{Q\Delta}{1\!-\!\Delta^2}\sin\phi,
\label{obj fun conic}
\end{equation}
where we have defined $\Delta=\eta_2-\eta_1$.  We readily see that the eccentricity of the ellipse is only a function of the priors. For equal priors, $\Delta=0$, the ellipse degenerates into the horizontal segment $v=Q$, $0\le u\le Q$, whereas for $Q=0$ it collapses into the origin $(u,v)=(0,0)$. As one increases $Q$, a family of similar ellipses is obtained. As they increase in size, their center moves up along the~$v$ axis. The line $u=v$ is the envelope of this family, as one can easily check using~Eq.~(\ref{obj fun conic}). The situation is illustrated in Fig.~\ref{fig:const z}~(a).

We now turn to Eq.~(\ref{zeta-means}). In terms of the new variables~$u$ and~$v$, this equation becomes the parabola
\begin{equation}
v=su+\frac{1\!-\!s^2}{2} - \frac{1\!-\!s^{2n}}{2s^{2n}}\left(\!u\!-\!s+\frac{\zeta}{\sqrt{1\!-\!s^{2n}}}\right)^2\kern-.2em.
\label{para}
\end{equation}

Because of the way we have written Eq.~(\ref{para}), it is apparent that for fixed input overlap $s$, the envelope of the family of parabolas obtained as $\zeta$ varies is the straight line given by the first two terms on the right hand side of Eq.~(\ref{para}), namely,
\begin{equation}
v=su+\frac{1-s^2}{2}.
\label{env par}
\end{equation}
As $\zeta$ increases from its minimum value, $\zeta=0$, the parabolas slide down along this straight line, Eq.~(\ref{env par}), without distortion. Any physically realizable cloner corresponds to a point $(u,v)$ that belongs to both, an ellipse and a parabola, for a given~$s$ and~$Q$. The optimal solution is given by the value of $\zeta$ that makes the parabola tangent to the ellipse. We see~that the upper end of the allowed $\zeta$ interval cannot exceed $\zeta_{\rm max}=s\sqrt{1-s^{2n}}-s^n\sqrt{1-s^2}$. This is the value of $\zeta$ for which the corresponding parabola contains the origin $(u,v)=(0,0)$, i.e., gives the solution for the deterministic cloner ($Q=0$). By increasing $n$ we make the parabolas narrower. In the limit $n\to\infty$, they become a vertical segment of height $(1+s^2)/2-s\zeta$ at the point~$u=s-\zeta$.

Ideally, we would like to find the optimal solution by computing the point of tangency between the conics we have just introduced. Unfortunately, this involves solving higher degree polynomial equation for which no formula for the roots exists. We therefore proceed as in~\cite{us2} and find the curve $F_{\rm FFR}(Q)$ in parametric form. The solution is (for details see appendixes)
\begin{eqnarray}
Q\!&=&\!{(1\!-\!\Delta^2)(1\!-\!s^2)\!-\!\gamma_n\left(\Delta\cot\phi\!+\!s\sqrt{1\!-\!\Delta^2}\right)^2\over2\left(1+\Delta\sin\phi-s\sqrt{1-\Delta^2}\cos\phi\right)},
\nonumber\\
\zeta_{\rm min} \!&=&\!{(1\!+\!\gamma_n)\sqrt{1\!-\!\Delta^2}s\!+\!\gamma_n\Delta\cot\phi\!-\!Q\cos\phi\over\sqrt{1+\gamma_n}\sqrt{1\!-\!\Delta^2}},
\label{sol par}
\end{eqnarray}
where $\gamma_n=s^{2n}/(1-s^{2n})$. We have $\lim_{n\to\infty}\gamma_n=0$. The upper end of the $\phi$ interval is given by the deterministic limit $Q=0$. This gives 
$\cot\phi_{\rm max}\!=\!-[s\!+\!\sqrt{(1\!-\!s^2)/\gamma_n}\,]\sqrt{1\!-\!\Delta^2}/\Delta$.
The lower end of the interval is determined by perfect cloning, i.e., $F_{\rm FFR}=1$, which in turn implies~$\zeta_{\rm min}=0$ and $s'=s^n$. 
However, no closed formula exists for~$\phi_{\rm min}$ and its value has to be computed numerically.  For smaller values of $\phi$, Eq.~(\ref{sol par}) does not give the optimal solution. These values would lead to failure probabilities larger than that required for perfect cloning. The strategy defined by Eq.~(\ref{sol par}) would produce separations below that required by perfect cloning, until full separation, $s'=0$ is attained. For this range of $Q$, the optimal scheme is perfect probabilistic cloning.

\begin{figure}[ht]
$$
\includegraphics[width=26em]{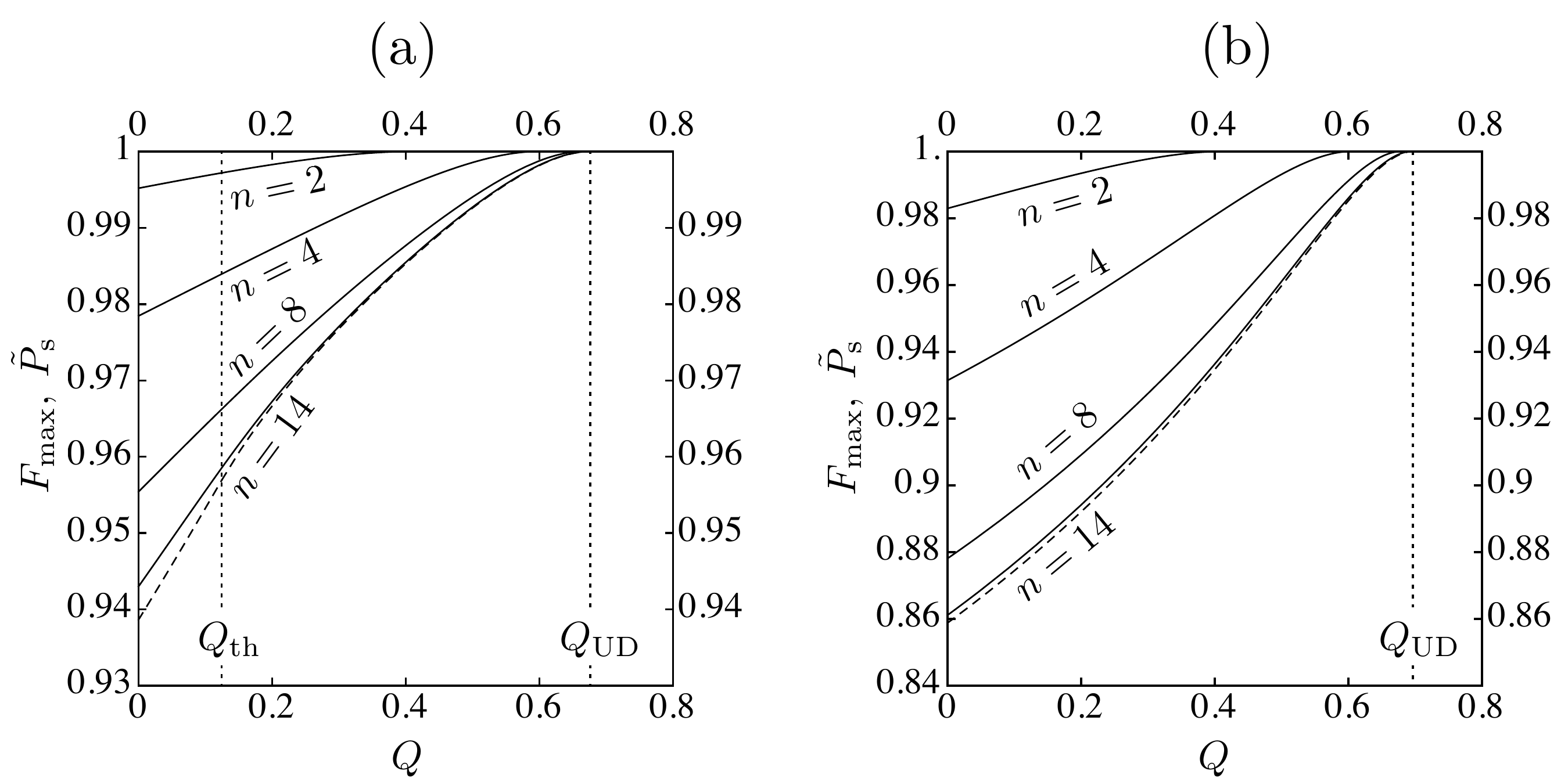}\quad
$$
\vspace{-2.5em}
\caption{ $F_{\rm FFR}$ vs. $Q$ for 
(a)~$s=0.8$, $\Delta=0.8$, and (b)~$s=0.7$, $\Delta=0.1$. Vertical (dotted) lines are drawn at the unambiguous failure rate $Q_{\rm UD}$ and at the threshold failure rate $Q_{\rm th}$, at which the FRIO scheme success probability, $\tilde P_{\rm s}$, (dashed line) changes regime. No such change occurs for the values of $s$ and~$\Delta$ in~(b). The lines attain the value $F_{\rm FFR}=1$ (perfect cloning) at $Q=Q_{\rm PC}$, given in Ref.~\cite{us1}. The lines approach the FRIO line, 
with discontinuous  second derivative at $Q=Q_{\rm th}$, as $n$ becomes larger. All curves have continuous derivatives for finite~$n$.}
\label{fig:tradeoff}
\end{figure}

Combining Eq.~(\ref{Fmax}) with Eq.~(\ref{sol par}) we obtain the tradeoff curve $F_{\rm FFR}(Q)$. Examples for different values of $n$ can be found in Fig.~\ref{fig:tradeoff}.

The limit $n \to \infty$ is of fundamental importance. We will show that the optimal protocol becomes ``measure and prepare". More precisely, the optimal cloning protocol can be implemented as a FRIO discrimination of the input states followed by a preparation of $|\psi^n_k\rangle$ if the discrimination is conclusive. If it is inconclusive, failure is reported. The fidelity for such protocol, conditioned on success (conclusive identification),  is 
\begin{equation}
F_{\rm FRIO}=\eta_1{p'_1+r'_1 s^{2n}\over\bar Q}+\eta_2{p'_2+r'_2 s^{2n}\over\bar Q} ,
\end{equation}
where $p'_k$ ($r'_k$) is the probability of (mis)identifying the input state $|\psi_k\rangle$. If $n\to\infty$, we readily see that $F^{\infty}_{\rm FRIO}=\tilde P_{\rm s}$, where~$\tilde P_{\rm s}=(\eta_1p'_1+\eta_2p'_2)/\bar Q$ is the FRIO average success probability conditioned on conclusive outcomes. 

Using the results  in~\cite{FRIO}, one can write
\begin{equation}
\tilde P_{\rm s}={\bar Q+\sqrt{\bar Q^2-(Q-Q_0)^2}\over2\bar Q},
\label{tildeP POVM}
\end{equation}
where 
$Q_0=2\sqrt{\eta_1\eta_2}s$ is the inconclusive (failure) probability for UD when $\eta_1\in[s^2/(1+s^2),1/2]$. For $\eta_1$ in this range, Eq.~(\ref{tildeP POVM}) holds for any physical value of the inconclusive probability $Q$, i.e., for $0\le Q\le Q_{\rm UD}=Q_0$. However, if the prior probabilities are very unbalanced, $\eta_1\in[0,s^2/(1+s^2)]$, two regimes exist. Eq.~(\ref{tildeP POVM}) holds only if $Q\le Q_{\rm th}$, where
\begin{equation}
Q_{\rm th}={2\eta_1\eta_2(1-s^2)\over 1-Q_0}.
\label{Q_th}
\end{equation}
For $Q_{\rm th}\le Q\le Q_{1}=\eta_1+\eta_2 s^2$, where $Q_{\rm UD}=Q_1$ in the above $\eta_1$ range, the three-outcome POVM cannot be implemented and the optimal measurement is projective (two-outcome). From the results in~\cite{FRIO} one can derive (see appendixes)
\begin{equation}
\tilde P_{\rm s}={\eta_2\over\bar Q}{(\eta_2-\eta_1)(\eta_2-Q)c^2+\bar Q s^2+2\eta_1 sc R\over1-4\eta_1\eta_2c^2},
\label{tildeP proj}
\end{equation}
where $c\!=\!\sqrt{1\!-\!s^2}$ and $R\!=\!\sqrt{Q\bar Q-\eta_1\eta_2 c^2}$, and check that the second derivative of $\tilde P_{\rm s}(Q)$ is discontinuous at~$Q_{\rm th}$.

Now that we have given the relevant FRIO results we come back to computing the asymptotic limit of our cloning scheme, $\lim_{n\to\infty}F_{\rm FFR}\equiv F_{\rm FFR}^\infty$.  To do that, we use our geometric picture. Since the parabolas in Eq.~(\ref{para}) become vertical segments in this limit, we notice that two different regimes will arise depending on whether the vertex of the ellipses in Eq.~(\ref{obj fun conic}) fall under the envelope (straight line) in Eq.~(\ref{env par}). The threshold is determined by the condition that the vertex ($\theta=0$) belongs to the envelope, i.e., satisfies Eq.~(\ref{env par}). We have $Q_{\rm th}/(1\!-\!\Delta^2)\!=\!s\, Q_{\rm th}/\sqrt{1\!-\!\Delta^2}+(1\!-\!s^2)/2$.
Solving for~$Q_{\rm th}$ we readily obtain Eq.~(\ref{Q_th}). For values of~$Q$ below the threshold, the corresponding ellipse and the vertical segment, located at $u=s-\zeta$, become tangent at the vertex, $\theta=0$, therefore $Q/\sqrt{1-\Delta^2}=s-\zeta_{\rm min}$. This equation can be written as 
$
2\sqrt{\eta_1\eta_2}\zeta_{\rm min}=Q_0-Q
$.
Substituting this into Eq.~(\ref{Fmax}) we obtain the expression on the right hand side of Eq.~(\ref{tildeP POVM}).  For~$Q_{\rm th}\le Q$, the ellipse and the straight segment cannot be tangent. The~ellipse merely touches the top of the vertical segment, so
\begin{eqnarray}
{Q\over\sqrt{1\!-\!\Delta^2}}\cos\phi\!&=&\!s\!-\!\zeta_{\rm min},\nonumber\\[-.2em]
{Q\over1\!-\!\Delta^2}\!+\!{Q\Delta\over1\!-\!\Delta^2}\sin\phi\!&=&\!{sQ\over\sqrt{1\!-\!\Delta^2}}\cos\phi+\!{1\!-\!s^2\over2}.\qquad
\end{eqnarray}
We can solve the second equation for $\cos\phi$ 
and substitute the result in the first equation to obtain
\begin{equation}
\zeta_{\rm min}={s[2(\Delta^2\!-\!Q)\!+\!(1\!+\! s^2)(1\!-\!\Delta^2)]\!-\!2\Delta c R\over2(s^2\!+\!\Delta^2 c^2)}.
\label{z_min arxiv}
\end{equation}
Substituting this in turn into Eq.~(\ref{Fmax}) we obtain (see appendixes) the expression on the right hand side of Eq.~(\ref{tildeP proj}). In summary, $F^{\infty}_{\rm FFR}=\tilde P_{\rm s}=F^\infty_{\rm FRIO}$ for all physical values of $Q$. Thus, in the limit of many copies optimal cloning can be implemented by FRIO discrimination followed by state preparation, as expected. 

In summary, we have provided the general solution to the long-standing problem of optimal cloning for two states with a fixed failure rate $Q$ (FFR cloning). The unequal prior case, $\eta_{1} \neq \eta_{2}$, uncovers remarkable phenomena that the very special equal prior case was unable to reveal. In particular, the convergence of cloning to FRIO discrimination as the number of clones becomes very large involves a discontinuity in the second derivative of the fidelity $F_{\rm FFR}(Q)$ at $Q_{\rm th}$, a phenomenon analogous to a second-order symmetry-breaking phase transition. Varying $Q$ between $0$ and $Q_{\rm PC}$ the fidelity of the clones varies from the fidelity of deterministic cloning to 1, i.e., perfect clones. Our geometric approach proved very powerful for both visualizing what the solution looks like qualitatively and for deriving the analytical solution. The same geometric approach can also be applied to other optimization problems that involve highly nonlinear constraints.

\begin{acknowledgments}
This publication was made possible through the support of a Grant from the John Templeton Foundation. The opinions expressed in this publication are those of the authors and do not necessarily reflect the views of the John Templeton Foundation. Partial financial support by a Grant from PSC-CUNY is also gratefully acknowledged. The research of EB was additionally supported by 
the Spanish MICINN, through contract FIS2013-40627-P, the Generalitat de
Catalunya CIRIT, contract  2014SGR-966, and ERDF: European Regional Development Fund. EB also thanks the hospitality of Hunter College during his research stay.
\end{acknowledgments}

\appendix

\section{Maximum Fidelity}

For the sake of completeness, we here derive the maximum (global) fidelity of the FFR cloner, Eqs.~(\ref{Fmax})  and~(\ref{zeta-means}). We follow Barnett and Chefles' derivation~\cite{Chefles+Barnett inter} with some modifications. We stick to our notation and recall that $|\Psi_k\rangle\in{\mathscr H}^{\otimes n}$ ($k=1,2$) is the state of the~$n$ approximate copies of $|\psi_k\rangle$, with prior probabilities~$\eta_k$. Then, the average fidelity conditioned on success~is
\begin{equation}
F=\tilde\eta_1|\langle \psi^n_1|\Psi_1\rangle|^2 +\tilde\eta_2|\langle \psi^n_2|\Psi_2\rangle|^2,
\end{equation}
where $\tilde\eta_k=\eta_k p_k/\bar Q$, $k=1,2$,  and $|\psi^n_k\rangle=|\psi_k\rangle^{\otimes n}$, i.e., $|\psi^n_k\rangle$ is the state of~$n$ perfect clones of $|\psi_k\rangle$. The optimal FFR cloner is that for which the average fidelity is maximum given a fixed failure probability $Q$.
\begin{figure}[b]
$$
\includegraphics[width=14em]{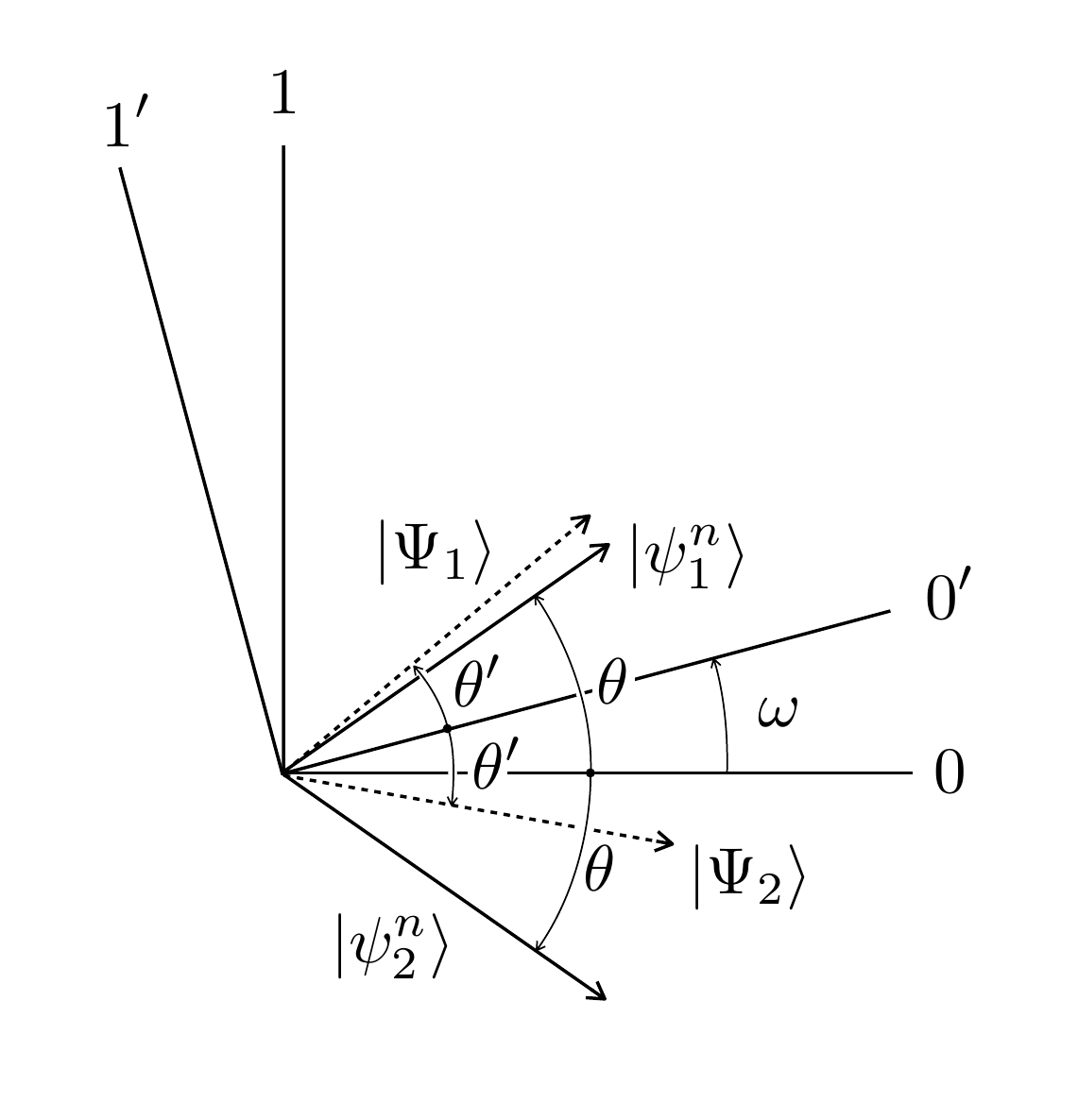}
$$
\caption{Choice of bases in Eqs.~(\ref{psis & Psis}) and (\ref{rot}).}
\label{fig:supp1}
\end{figure}

With no loss of generality we may write
\begin{eqnarray}
|\psi^n_k\rangle&=&\cos\theta\,|0\rangle-(-1)^k\sin\theta\,|1\rangle,\nonumber\\
|\Psi_k\rangle&=&\cos\theta'\,|0'\rangle-(-1)^k\sin\theta'\,|1'\rangle,
\label{psis & Psis}
\end{eqnarray}
where $\{|0\rangle,|1\rangle\}$ and $\{|0'\rangle,|1'\rangle\}$ are two conveniently chosen orthogonal bases. Then, $0\le\theta\le\pi/4$, $0\le\theta'\le\pi/4$, and a simple calculation leads to
\begin{eqnarray}
s^n&\equiv&|\langle\psi_1|\psi_2\rangle|^n=|\langle\psi^n_1|\psi^n_2\rangle|=\cos2\theta,\nonumber\\[.5em]
s'&\equiv&|\langle\Psi_1|\Psi_2\rangle|=\cos2\theta'.
\label{eb ths = ss}
\end{eqnarray}
Since the two orthonormal bases $\{|0\rangle,|1\rangle\}$ and $\{|0'\rangle,|1'\rangle\}$ must be connected by a unitary, we can write
\begin{eqnarray}
|0'\rangle&=&\cos\omega\,|0\rangle-\sin\omega\,|1\rangle,\nonumber\\
|1'\rangle&=&\sin\omega\,|0\rangle+\cos\omega\,|1\rangle  .
\label{rot}
\end{eqnarray}
The angle $\omega$ is a free parameter.  It gives us the orientation of the basis $\{|0'\rangle,|1'\rangle\}$ relative to~$\{|0\rangle,|1\rangle\}$. The geometry behind this choice of bases and parameters is sketched in Fig.~\ref{fig:supp1}. 
Our aim is to find the value of~$\omega$ that maximizes the fidelity~$F$.

Using the definitions above we can write
\begin{eqnarray}
|\Psi_k\rangle&=&\left[\cos\theta'\,\cos\omega-(-1)^k\sin\theta'\sin\omega\right]|0\rangle\nonumber\\
&-&\left[\cos\theta'\,\sin\omega+(-1)^k\sin\theta'\cos\omega\right]|1\rangle ,
\end{eqnarray}
which can be simplified as
\begin{equation}
|\Psi_{\!k}\rangle\!\!=\!\cos\!\left[\theta'\!\!+\!(\!-1)^k\omega\right]\!|0\rangle\!-\!(\!-1)^k\!
\sin\!\left[\theta'\!\!+\!(\!-1)^k\omega\right]\!|1\rangle.
\end{equation}
We can now easily compute the overlaps between perfect and imperfect clones,
\begin{equation}
\langle\psi^n_k|\Psi_k\rangle=\cos\left[\theta-\theta'-(-1)^k\omega\right] .
\end{equation}
Substituting in  the definition of $F$ we have
\begin{equation}
F\!=\!\tilde\eta_1\cos^2\left(\theta\!-\!\theta'\!\!+\!\omega\right)+\tilde\eta_2\cos^2\left(\theta\!-\!\theta'\!\!-\!\omega\right) .
\end{equation}
This expression can be written more conveniently as
\begin{equation}
F\!=\!{1\over2}\!+\!{\cos\!\left(2\theta\!-\!2\theta'\right)\!\cos2\omega
\!+\!\tilde\Delta\sin\!\left(2\theta\!-\!2\theta'\right)\!\sin2\omega
\over 2},
\end{equation}
where we have defined $\tilde\Delta=\tilde\eta_2-\tilde\eta_1$.
To optimize, we apply Schwarz inequality to the last two terms and recall that the inequality is saturated if
\begin{eqnarray}
\sin2\omega&=&\lambda\,\tilde\Delta\sin(2\theta\!-\!2\theta'),\nonumber\\
\cos2\omega&=&\lambda\cos(2\theta\!-\!2\theta'),
\end{eqnarray}
for some real number $\lambda$.
This leads us immediately to the equations
\begin{eqnarray}
\sin\!2\omega&\!=\!&
{\tilde\Delta\sin(2\theta-2\theta')\over\sqrt{\!\cos^2(2\theta\!-\!2\theta')\!+\!\tilde\Delta^2\sin^2(2\theta\!-\!2\theta')}},\nonumber\\[.5em]
\cos\!2\omega&\!=\!&
{\cos(2\theta-2\theta')\over\sqrt{\!\cos^2(2\theta\!-\!2\theta')\!+\!\tilde\Delta^2\sin^2(2\theta\!-\!2\theta')}},
\end{eqnarray}
which in turn imply
\begin{equation}
F_{\rm max}(\theta')\!=\!{1\over2}+{1\over2}
\sqrt{\!\cos^2(2\theta\!-\!2\theta')\!+\!\tilde\Delta^2\sin^2(2\theta\!-\!2\theta')} .
\label{F_max_1}
\end{equation}
We note in passing that for $\tilde\eta_1=\tilde\eta_2$ we have~$\omega=0$, as expected.
Eq.~(\ref{F_max_1}) can be further simplified to give 
\begin{equation}
F_{\rm max}(\theta')={1\over2}+{1\over2}
\sqrt{\!1-\!4\tilde\eta_1\tilde\eta_2\sin^2(2\theta\!-\!2\theta')} .
\vspace{.2em}
\label{F_max}
\end{equation}

Recalling the definition of $\tilde\eta_k$, we can write
\begin{equation}
F_{\rm max}(\theta')=\frac{\bar Q+\sqrt{\bar Q^2-4\eta_1\eta_2\zeta^2}}{2\bar Q} ,
\label{Fmax_supp}
\end{equation}
where we have defined $\zeta\ge0$ as
\begin{equation}
\zeta=\sqrt{p_1p_2}\sin(2\theta-2\theta').
\end{equation}
We can ged rid of trigonometric functions and $s'$ in the definition of~$\zeta$ by using Eqs.~(\ref{eb ths = ss}) and the unitarity constraint, $s=\sqrt{p_1p_2}\,s'+\sqrt{q_1q_2}$, Eq.~(3). It yields
\begin{eqnarray}
\zeta&=&(s-\sqrt{q_1q_2})\sqrt{1-s^{2n}}\nonumber\\
&-&s^n\sqrt{1-s^2+2s\sqrt{q_1q_2}-(q_1+q_2)}.
\label{zeta-means_supp}
\end{eqnarray}
This is Eq.~(\ref{zeta-means}).
The maximum value of~$F_{\rm max}(\theta')$ 
is denoted by $F_{\rm FFR}$. It is given by the minimum positive value of $\zeta$, which leads to Eqs.~(\ref{Fmax}).

\section{\boldmath $\zeta_{\rm min}(Q)$ in parametric form} 

In the previous section we have shown that for a given failure rate $Q$ the maximum fidelity is given by the minimum positive value of~$\zeta$ (which we denote with $\zeta_{\rm min}$) through Eq.~(\ref{Fmax_supp}). As explained in the main text, to obtain $\zeta_{\rm min}(Q)$ we need to find the value of $\zeta$ for which the parabola
 \begin{equation}
 v=su+\frac{1\!-\!s^2}{2} - \frac{1\!-\!s^{2n}}{2s^{2n}}\left(\!u\!-\!s+\frac{\zeta}{\sqrt{1\!-\!s^{2n}}}\right)^2
\label{para_supp}
 \end{equation}
becomes tangent to the ellipse
\begin{equation}
u\!=\!\frac{Q}{\sqrt{1\!-\!\Delta^2}}\cos\phi,\quad
v\!=\!\frac{Q}{1\!-\!\Delta^2}\!+\!\frac{Q\Delta}{1\!-\!\Delta^2}\sin\phi.
\label{obj fun conic_supp}
\end{equation}
Deriving an explicit expression for $\zeta_{\rm min}(Q)$ involves finding the zeroes of higher degree polynomial equations, for which no formula is known. So, instead, we here derive a parametric expression for $\zeta_{\rm min}(Q)$.

At the tangency point, $(u,v)$ must satisfy both Eq.~(\ref{para_supp}) and Eq.~(\ref{obj fun conic_supp}) and their derivatives $dv/du$ must be equal. From Eq.~(\ref{para_supp}) we readily see that
 \begin{equation}
\!u\!-\!s+\frac{\zeta}{\sqrt{1\!-\!s^{2n}}}=\gamma_n\left(s- {dv\over du}\right),
\label{der para}
 \end{equation}
where we have defined $\gamma_n\equiv s^{2n}/(1-s^{2n})$ [equivalently, $s^{2n}=\gamma_n/(1+\gamma_n)$]. From Eq.~(\ref{obj fun conic_supp}), we have
\begin{equation}
{dv\over du}={dv/d\phi\over du/d\phi}=-{\Delta\over\sqrt{1-\Delta^2}}\cot\phi.
\label{der obj fun conic}
\end{equation}
We next substitute Eq.~(\ref{der obj fun conic}) into Eq.~(\ref{der para}) and substitute the resulting expression onto the right hand side of Eq.~(\ref{para_supp}) to obtain
\begin{equation}
v=su+\frac{1\!-\!s^2}{2} - \frac{\gamma_n}{2}\left(\!s\!+{\Delta \cot\phi\over\sqrt{1-\Delta^2}}\right)^2.
\label{tang}
\end{equation}
Finally, we substitute the expressions of $u$ and $v$ given in Eq.~(\ref{obj fun conic_supp}) into Eq.~(\ref{tang}) and solve for $Q$ to obtain
\begin{equation}
Q={(1\!-\!\Delta^2)(1\!-\!s^2)\!-\!\gamma_n\left(\Delta\cot\phi\!+\!s\sqrt{1\!-\!\Delta^2}\right)^2\over2\left(1+\Delta\sin\phi-s\sqrt{1-\Delta^2}\cos\phi\right)}.
\end{equation}
The analogous expression for $\zeta_{\rm min}$ is easily obtained by solving  for $\zeta$ in Eq.~(\ref{der para}) and using again Eq.~(\ref{obj fun conic_supp}) and Eq.~(\ref{der obj fun conic}). The result is
\begin{equation}
\zeta_{\rm min}={(1\!+\!\gamma_n)\sqrt{1\!-\!\Delta^2}s\!+\!\gamma_n\Delta\cot\phi\!-\!Q\cos\phi\over\sqrt{1+\gamma_n}\sqrt{1\!-\!\Delta^2}} .
\end{equation}
These last two equations are collected in Eq.~(\ref{sol par}).

\section{\boldmath Derivation of $\tilde P_{\rm s}$ for $Q\ge Q_{\rm th}$.} 
We next derive Eq.~(\ref{tildeP proj}).
Using our current notation and defining $c=\sqrt{1-s^2}$, Eq.~(15) in~\cite{FRIO}~is
\begin{equation}
\tilde P_{\rm s}=\eta_2\left(\sqrt{1-\tilde P_{\rm s}\over\eta_1}\,s-\sqrt{{1\over\bar Q}-{1-\tilde P_{\rm s}\over\eta_1}}\,c\right)^2,
\end{equation}
as the error probability is $P_{\rm e}=\bar Q-P_{\rm s}=\bar Q(1-\tilde P_{\rm s})$ and we are assuming that $\eta_1\le\eta_2$. Solving for $\tilde P_{\rm s}$, and after some algebra, we obtain
\begin{equation}
\tilde P_{\rm s}={\eta_2\over\bar Q}{(\eta_2-\eta_1)(\eta_2-Q)c^2+\bar Q s^2+2\eta_1 sc R\over1-4\eta_1\eta_2c^2},
\label{tildeP proj_supp}
\end{equation}
where
\begin{equation}
R=\sqrt{Q\bar Q-\eta_1\eta_2 c^2}. 
\label{def R}
\end{equation}
\vspace{.1em}

\section{\boldmath Derivation of $F^\infty_{\rm FFR}$ for $Q\ge Q_{\rm th}$.} 

We first show that the maximum value of the argument of the square root in 
Eq.~(\ref{Fmax}) becomes a perfect square. More precisely, 
\begin{eqnarray}
\hspace{-1em}\bar Q^2\!\!-\!4\eta_1\eta_2\zeta^2_{\rm min}\!&=&\!{1\over  4(s^2\!+\!\Delta^2 c^2)^2}\!\left\{
2sc(1\!-\!\Delta^2)R\right.\nonumber\\
\!&+&\!\!
\left.
\Delta\!
\left[
2(\Delta^2\!-\!Q)\!+\!(1\!+\!s^2)(1\!-\!\Delta^2)
\right]
\right\}^{\!2}\!\!.
\label{perfect square}
\end{eqnarray}
To show this, rather than dealing with radicals, which usually involves cumbersome algebraic manipulations, we look at $R$ as an independent  variable. Recalling the expression for $\zeta_{\rm min}$ in Eq.~(\ref{z_min arxiv}), we can write
\begin{equation}
\bar Q^2\!-\!4\eta_1\eta_2\zeta^2_{\rm min}\!=\!{A\!+\!BR\!+\!CR^2\over4(s^2\!+\!\Delta^2 c^2)^2},
\label{eq ABC}
\end{equation}
where
\begin{eqnarray}
A\!&=&\!
4(s^2\!+\!\Delta^2c^2)^2\bar Q^2\!-\!
s^2(1\!-\!\Delta^2)\nonumber\\[.5em]
\!&\times&\!\!
\left[
2(\Delta^2\!-\!Q)\!+\!(1\!+\!s^2)(1\!-\!\Delta^2)
\right]^2,
\nonumber\\[1em]
B\!&=&\!
4sc\,\Delta(1\!-\!\Delta^2)\!
\left[
2(\Delta^2\!\!-\!Q)\!+\!(1\!+\!s^2)(1\!-\!\Delta^2)
\right],\nonumber
\\[.5em]
C\!&=&\!-4c^2\Delta^2(1\!-\!\Delta^2),
\label{ABC}
\end{eqnarray}
and we have used that $4\eta_1\eta_2=1-\Delta^2$.
We next note that, by definition of $R$, Eq.~(\ref{def R}),
\begin{equation}
4c^2(1\!-\!\Delta^2)(s^2\!+\!\Delta^2c^2)\!
\left(\!
R^2\!\!-\!Q\bar Q\!+\!c^2{1\!-\!\Delta^2\over4}
\right)\!
\!=\!0.
\end{equation}
So, 
we can make the replacements
\begin{eqnarray}
A\!&\to&\!A\!-\!4c^2(1\!-\!\Delta^2)(s^2+\Delta^2c^2)\!
\!\left(\!\!
Q\bar Q\!-\!c^2{1\!-\!\Delta^2\over4}\!
\right),
\nonumber\\[.5em]
C\!&\to&\!C\!+\!4c^2(1-\Delta^2)(s^2+\Delta^2c^2),
\end{eqnarray}
without altering the equality in Eq.~(\ref{eq ABC}).
The resulting expression can be easily seen to be the right hand side of~Eq.~(\ref{perfect square}). By using this equation in Eq.~(\ref{Fmax}) it is straightforward to obtain~Eq.~(\ref{tildeP proj}).

\end{document}